# Sub-20-Attosecond Timing Jitter Mode-Locked Fiber Lasers

Hyoji Kim, Peng Qin, Youjian Song, Heewon Yang, Junho Shin, Chur Kim, Kwangyun Jung, Chingyue Wang, and Jungwon Kim

*Abstract*— We demonstrate 14.3-attosecond timing jitter [integrated from 10 kHz to 94 MHz offset frequency] optical pulse trains from 188-MHz repetition-rate mode-locked Yb-fiber lasers. In order to minimize the timing jitter, we shorten the non-gain fiber length to shorten the pulsewidth and reduce excessive higher-order nonlinearity and nonlinear chirp in the fiber laser. The measured jitter spectrum is limited by the amplified spontaneous emission limited quantum noise in the 100 kHz – 1 MHz offset frequency range, while it was limited by the relative intensity noise-converted jitter in the lower offset frequency range. This intrinsically low timing jitter enables sub-100-attosecond synchronization between the two mode-locked Yb-fiber lasers over the full Nyquist frequency with a modest 10-kHz locking bandwidth. The demonstrated performance is the lowest timing jitter measured from any free-running mode-locked fiber lasers, comparable to the performance of the lowest-jitter Ti:sapphire solid-state lasers.

## I. INTRODUCTION

RECENTLY there have been remarkable progresses in the characterization and optimization of timing jitter in femtosecond passively mode-locked lasers. Currently, various types of passively mode-locked solid-state and fiber lasers can generate optical pulse trains with high-frequency timing jitter in the sub-femtosecond regime [1-7]. This rapid advancement in timing jitter optimization in the last few years was possible due to well-established noise models of mode-locked lasers [8-11] and the invention of attosecond-resolution timing detection methods [12,13]. The availability of femtosecond mode-locked lasers with sub-femtosecond timing jitter can significantly improve many high-precision scientific and engineering applications such as sub-cycle optical pulse synthesis [14,15], ultralow phase-noise microwave signal generation [16-18], high-resolution and high-speed photonic

Submitted on November 26, 2013. This research was supported by the National Research Foundation (NRF) of South Korea (Grant numbers 2012R1A2A01005544 and 2013M1A3A02042273).

Hyoji Kim, Heewon Yang, Junho Shin, Chur Kim, Kwangyun Jung, and Jungwon Kim are with the School of Mechanical, Aerospace and Systems Engineering, Korea Advanced Institute of Science and Technology (KAIST), Daejeon 305-701, South Korea (e-mail: jungwon.kim@kaist.ac.kr).

Peng Qin, Youjian Song, and Chingyue Wang are with the College of Precision Instruments and Optoelectronics Engineering, Tianjin University, Tianjin 300072, China.

analog-to-digital converters [19,20], down-conversion receivers [21], and local and remote synchronization for large-scale scientific facilities (such as free-electron lasers and ultrafast electron diffraction sources) and ultrafast pump-probe experiments [22]. In particular, for attosecond synchronization between different mode-locked lasers and/or microwave sources, high-frequency timing jitter of mode-locked lasers should be minimized, because the free-running high-frequency timing jitter outside the locking bandwidth will ultimately limit the achievable synchronization performance. Therefore, for the advancement of the attosecond-precision ultrafast photonics [23], it would be highly desirable to scale down the timing jitter of free-running mode-locked laser oscillators to the lowest possible level, well into the few-attosecond regime.

So far, the lowest timing jitter measured from mode-locked lasers was demonstrated from a sub-10-fs, Kerr-lens mode-locked (KLM), solid-state Ti:sapphire laser, which was measured to be ~10 as (from 10 kHz to 41.5 MHz integration bandwidth) [3]. This was possible due to much shorter pulsewidth (~10 fs), higher intra-cavity pulse energy and less higher-order nonlinearity and chirp, compared to mode-locked fiber lasers. If a comparable jitter performance can be achieved from mode-locked fiber lasers, it would be highly desirable due to their advantages of more compact size, lower cost, less alignment sensitivity and better robustness, which are essential for applications outside laboratory environments. The best jitter performance from mode-locked fiber lasers has been so far about 100-as regime: 175-as from Yb-fiber lasers [1] and 70-as from Er-fiber lasers [2] for integration bandwidth of 10 kHz – 40 MHz (Nyquist frequency). This performance was obtained by setting the intra-cavity dispersion close to zero (e.g., within -0.004 to 0 $ps^2$ range) in the stretched-pulse mode-locking regime: both timing jitter directly coupled from amplified spontaneous emission (ASE) noise and timing jitter indirectly coupled from center frequency fluctuation via intra-cavity dispersion can be minimized in this condition. Further reduction of jitter might have been limited by the higher-order nonlinearity and large nonlinear chirp in the fiber lasers. As shown in [2], although the jitter was optimized to the sub-100-as regime, the measured jitter level was still several dB higher than the jitter level predicted by the analytic noise theory of stretched-pulse mode-locked lasers [8].

In this paper, we show the most recent achievement in timing jitter performance of mode-locked fiber lasers. By reducing the non-gain fiber length in the fiber laser, the timing jitter of mode-locked Yb-fiber laser can be scaled down to the

sub-20-as regime when integrated from 10 kHz to Nyquist frequency (94 MHz in this work). This result is about a factor of 10 improvement from the previous lowest-jitter Yb-fiber laser result [1], and is now comparable to the performance of the lowest timing jitter Ti:sapphire lasers [3]. The demonstrated result shows the promise of mode-locked fiber lasers as ultralow-jitter optical pulse train sources for future attosecond-precision ultrafast photonic applications.

## II. LASER DESIGN AND EXPERIMENTAL SETUP

### A. Laser Design and Implementation

The mode-locked fiber laser we investigated in this work is a nonlinear polarization rotation (NPR)-based passively mode-locked Yb-fiber laser, shown in Fig. 1(a). It is basically a standard NPR-based, stretched-pulse fiber laser structure: the change here is that we made the non-gain fiber section as short as possible to the level allowed by the splicing machine we used. As a result, compared to the previous 175-as jitter, 80-MHz Yb-fiber laser in [1], the non-gain fiber length is reduced from 2 m to 43.5 cm. Recently this short non-gain fiber approach has also been used to reduce the pulsewidth from mode-locked Er- and Yb-fiber lasers [24,25] by reducing the nonlinear pulse evolution in the oscillator. This shorter average pulsewidth in the oscillator is advantageous for reducing the timing jitter as well, because the integrated timing jitter directly coupled by the ASE noise is proportional to the average pulsewidth [8,10]. In addition, shorter non-gain fiber length reduces excessive higher-order nonlinearity and nonlinear chirp accumulated in the long non-gain fiber section. Especially, we tried to keep the non-gain fiber section after the Yb-gain fiber as short as possible (11 cm), which helps mitigating large nonlinearity experienced by a high peak power pulse after the amplification.

The overall idea of reducing non-gain fiber length is to make the fiber laser closer to the solid-state laser. Although the fiber laser with free-space section still has an alignment sensitivity compared to all-fiber ring structure, due to the use of a WDM coupler for pumping the gain fiber, it is much less alignment-sensitive compared to the solid-state lasers (especially KLM lasers) and is more robust and long-term stable. When we find the right mode-locked condition, our fiber lasers operated without significant change in power and mode-locked condition for more than several weeks. Therefore, we believe that the short non-gain fiber approach can combine the advantages of fiber lasers and solid-state lasers for ultralow-jitter photonic sources: it can have better robustness, easier implementation and lower cost of fiber lasers while having similar jitter performance of the best KLM Ti:sapphire lasers.

More detailed information on the constructed laser is following. The used gain fiber is 22 cm of CorActive Yb214 fiber, which is the same length as the laser in [1]. The free-space section was also built as compact as possible with total length of 62 cm. As a result of reduced non-gain fiber length and free-space section length, the repetition rate of the laser is increased to 188 MHz (from 80 MHz in [1]). When the pump power of 570 mW is used, and the resulting output power and the intra-cavity pulse energy are 210 mW and 2.2 nJ, respectively. In order to make the net intra-cavity dispersion at close-to-zero, we used a 600 lines/mm grating pair and set the intra-cavity dispersion within -0.002 ps² to 0 ps² range at 1040 nm center wavelength. Note that the beam height of the input and output pulses at the grating pair is adjusted by using a right-angle mirror pair as an end mirror. Fig. 1(b) shows the measured optical spectrum of the fiber laser. After dechirping the outputs by a 300 lines/mm grating pair, we measured the pulsewidth with an auto-correlator, which shows 40 fs full-width half maximum (FWHM) pulsewidth. This pulsewidth is significantly shorter than the pulsewidth of an 80-MHz laser in [1] (~60 fs). Fig. 1(c) shows the measured relative intensity noise (RIN) of the fiber laser. The integrated rms RIN is 0.015 % when integrated from 100 Hz to 1 MHz offset frequency. Note that the measured RIN performance is among the lowest RIN levels reported so far from mode-locked fiber laser oscillators [26,27].

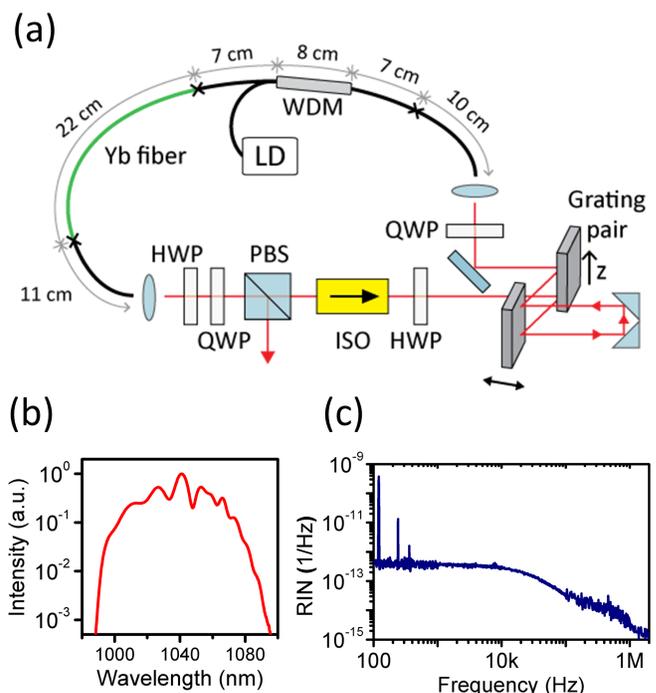

Fig. 1. (a) Structure of the mode-locked Yb-fiber laser used for minimizing timing jitter. HWP, half-wave plate; ISO, isolator; LD, 600 mW, 976 nm single-mode pump diode; QWP, quarter-wave plate; WDM, 1030 nm-976 nm wavelength division multiplexing coupler. Note that the used WDM coupler is a reflective type one. (b) Measured optical spectrum. (c) Measured relative intensity noise (RIN) spectrum.

### B. Experimental Setup for Timing Jitter Measurement

Figure 2(a) shows the timing jitter measurement experimental setup. For the timing jitter measurement of free-running mode-locked lasers using the timing detector method [12], two independent and almost identical lasers are required. Thus, in addition to a ring laser (as shown in Fig. 1(a)), we also built a sigma-cavity fiber laser with same fiber lengths, with the only difference of having a piezoelectric transducer (PZT)-mounted mirror in the free-space section for the repetition rate locking. Both lasers are set to have almost identical conditions and parameters (e.g., the length of fiber sections, output power, optical spectra) and are different only in the free-space section configuration. After dechirped by 300 lines/mm grating pairs, the optical pulse trains from the two lasers are combined by the polarization beam splitter (PBS in Fig. 2(a)) applied to the balanced optical cross-correlator (BOC) with perpendicular polarization to each other. More general information on the operation of BOC as a timing detector can be found in refs. [3,12,23]. In this work, in order to improve the timing detection sensitivity, we modified the BOC structure. The BOC in this work consists of type-II phase matched beta-barium borate (BBO) crystals with thickness of 400 μm and 800 μm. The BBO1 and BBO3 in Fig. 2(a) are used for the second-harmonic generation (SHG) for timing detection. The BBO2 in Fig. 2(a) is used for finely tuning the delay offset between the two timing signals (i.e., the time separation between two peaks in Fig. 2(b)) by tuning birefringence between the two polarization components of input pulses. With the optimization of the BOC, we could improve the timing detection sensitivity by a factor of 10 dB compared to the result in [1].

The timing jitter measurement is performed at the zero crossing of the BOC output with the low-bandwidth laser synchronization. In order to evaluate the shot-noise limited detection noise floor level and also to check whether the zero crossing of the BOC output happens at the middle point of the single-arm cross-correlation, we evaluated the SHG power of each photodiode when the BOC is balanced. We could confirm that the zero crossing of the BOC output happens around the intended middle point (0.46 V) of individual SHG signals where the detection sensitivity is the maximum. The shot noise limited jitter spectral density level of the BOC output (for a single laser) can be obtained by $S_{\Delta t2}(f) = 2eVG/k^2$ (fs²/Hz), where $e$ is the electron charge (1.6×10⁻¹⁹ C), $V$ is the voltage output contributed from a single photodiode at the balanced point (0.46 V), $G$ is the trans-impedance gain of the used balanced photodetector (3 kV/A), and $k$ is the detection slope of the BOC (31.9 mV/fs). The resulting shot noise limit is ~4.3×10⁻¹³ fs²/Hz, which fits fairly well with the measurement results shown in Figs. 4 and 5 in Section III. This corresponds to the unprecedented 6.4-as resolution integrated over 94-MHz bandwidth, which is the Nyquist frequency of the 188-MHz Yb-fiber lasers.

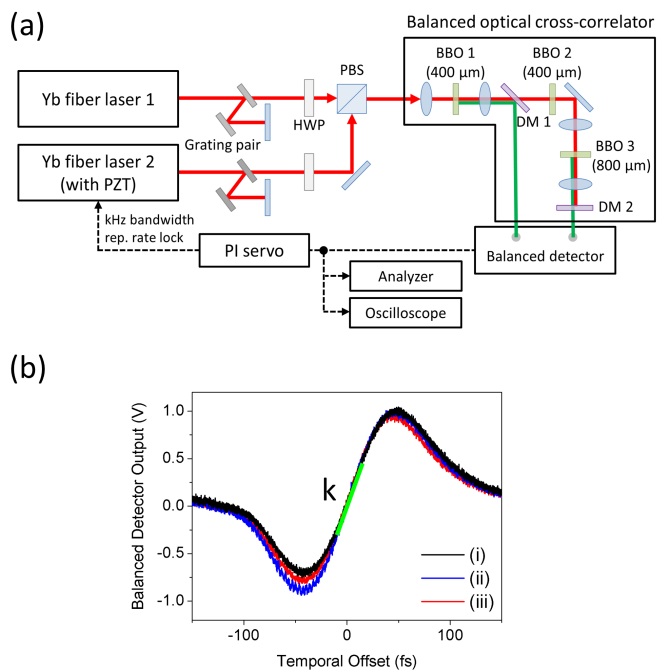

Fig. 2. (a) Experimental setup for timing jitter measurement. DM, dichroic mirror; HWP, half wave plate; PBS, polarization beam splitter; PZT, piezoelectric transducer. (b) Measured timing detection signal from the BOC. Repetition rate detuning of (i) 20 Hz, (ii) 40 Hz and (iii) 50 Hz.

Fig. 2(b) shows the measured timing detection signals from the BOC when the repetition rates of two lasers are slightly detuned. To confirm the validity of the slope determination, we measured the BOC outputs when the repetition rate detuning is set to 20 Hz, 40 Hz, and 50 Hz. As shown in Fig. 2(b), the BOC output and the timing detection slope ($k$ in Fig. 2(b)) is consistent regardless of the detuning amount, which shows the validity of the timing detection slope determination process. With ten independent measurements, the timing detection slope is determined to be $k = 31.9$ (±0.8) mV/fs.

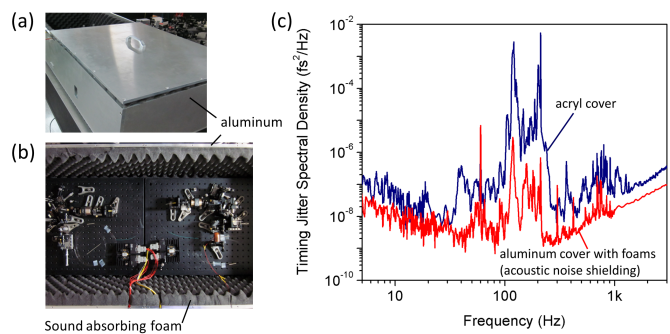

Fig. 3. Acoustic noise shielding. (a) Aluminum case. (b) Inside the aluminum case with sound-absorbing foams attached. (c) Timing jitter spectral density when an acryl case is used (blue curve) and when an aluminum case with sound-absorbing foam is used. The timing jitter density from the acoustic noise coupling is reduced by up to 40 dB.

Another technical improvement for jitter measurement of ultralow-jitter fiber lasers is the use of better acoustic noise shielding. In order to lower the necessary locking bandwidth and, as a result, to extend the measurable free-running timing jitter spectral density to lower offset frequency range, the technical noise (acoustic noise in particular) in the low offset frequency should be minimized. Fig. 3(a) and 3(b) show the photos of the constructed acoustic-noise shielded aluminum case. Previously, we only used an acryl cover and suffered from strong acoustic noise caused by the fan noise from instruments in the laboratory, which limited the measurement locking bandwidth and also made the synchronization more difficult to obtain. To circumvent this problem, a new aluminum laser case with sound absorbing foam attached inside is used. Fig 3(c) compares the measured timing jitter spectral density when an acryl cover is used (blue curve) and when a new aluminum cover with foams is used (red curve). Note that the locking bandwidth is set to ~10 kHz and the loop gain is set to maximally allowed level. One can see that the acoustic noise peaks ranging from 30 Hz to 300 Hz are greatly reduced by up to 40 dB. As a result, the in-loop timing jitter contribution from the acoustic noise peaks is reduced from 139 as to 5 as in the 30 Hz – 300 Hz range. With the acoustic noise shielding, the lowest locking bandwidth we could obtain was reduced to ~2 kHz. This means that, above ~2 kHz offset frequency, the timing jitter spectral density of free-running mode-locked lasers can be measured, as will be shown in Section III.

## III. Timing Jitter Measurement Results and Analysis

### A. Measurement Results

Using the BOC-based timing jitter measurement system explained in Section II.B, we could characterize the timing jitter spectral density of the free-running mode-locked Yb-fiber lasers above ~2 kHz offset frequency. The measured power spectral density from the BOC output is divided by two to reflect the jitter contribution from one laser, assuming the two lasers are uncorrelated and almost identical. By adjusting the intra-cavity wave plates in the fiber laser oscillators, we searched for the minimal timing jitter condition. Fig. 4 shows the measured minimal timing jitter spectral density data for a single laser with 7 different locking conditions (in terms of locking bandwidth, phase margin at the locking bandwidth, and loop gain of the phase-locked loop (PLL)). Note that these measurements were taken over 4 months on January 8, 16, 21, 24, 26 (twice), and April 28, 2013, and could be obtained consistently by finding the right mode-locking conditions for the minimal timing jitter. Regardless of the locking bandwidth and the PLL loop gain, we can find the timing jitter spectral density data above 10 kHz offset frequency are almost identical, which shows that the jitter measurement is indeed limited by the free-running laser itself and not distorted by the repetition rate locking PLL operation. Above ~1 MHz offset frequency, the measured jitter spectrum is limited by the shot-noise-limited BOC resolution, as explained in Section II.B, which is at ~4.3×10⁻¹³ fs²/Hz. The minimal integrated rms timing jitter from 10 kHz to 94 MHz (Nyquist frequency) is 14.3 (±0.4) as, where the stated error is determined by the uncertainty in the BOC timing slope measurements. Also note that the shot-noise-limited BOC noise floor already contributes 6.8-as in the measured jitter. The measured result is more than a factor of 10 lower than the previous timing jitter record of mode-locked Yb-fiber lasers [1] and, to our knowledge, corresponds to the lowest high-frequency timing jitter measured from any free-running mode-locked fiber lasers.

This intrinsically low timing jitter of mode-locked fiber lasers enables attosecond-level active synchronization between the two fiber lasers. Fig. 5 shows the best synchronization performance between the two Yb-fiber lasers we could obtain. The residual rms timing jitter between the two fiber lasers is 93 as integrated from 10 Hz to the Nyquist frequency (94 MHz) when a modest locking bandwidth of ~10 kHz is used, limited by the used PZT-mounted mirror. Note that PZT-mounted mirrors with locking bandwidth over 100 kHz [28] and even more broadband electro-optic modulators [29] are available for the repetition rate locking. By extending the locking bandwidth to >100 kHz, we can anticipate that the synchronization performance between mode-locked fiber lasers can be further optimized to the sub-20-attosecond regime over the full Nyquist frequency.

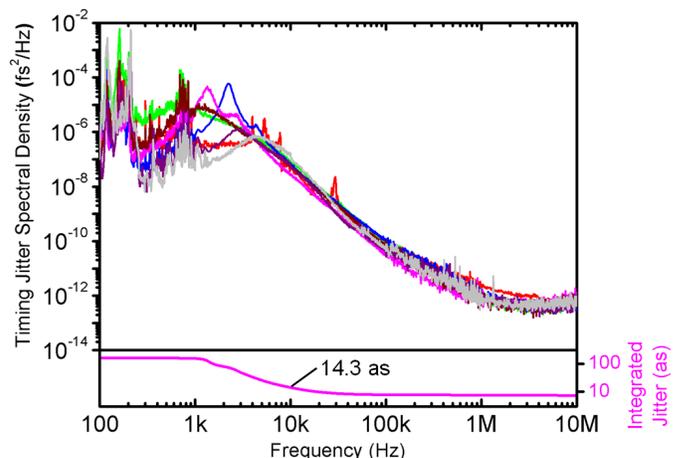

Fig. 4. Timing jitter spectral density measurement results with 7 different locking bandwidth and loop gain conditions for low-bandwidth (<10 kHz) laser synchronization. Regardless of the locking conditions the measurement dates, the measured timing jitter data above 10 kHz offset frequency are consistent and represent the timing jitter of a free-running laser oscillator. The lowest measured timing jitter is 14.3 as (rms) integrated from 10 kHz to 94 MHz offset frequency.

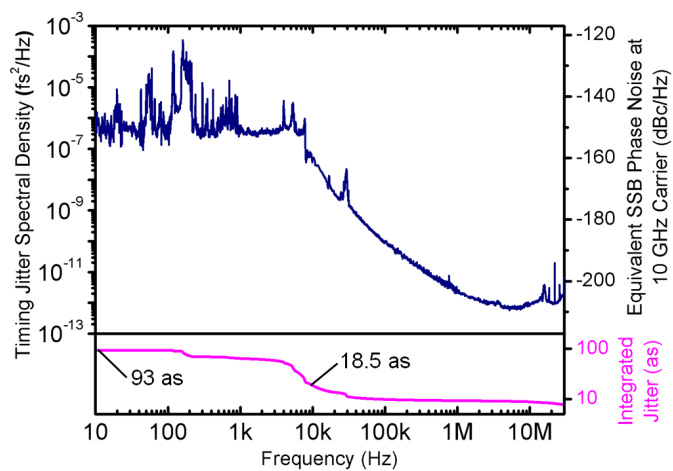

Fig. 5. Synchronization between the two 188-MHz Yb-fiber lasers. The residual timing jitter between the two lasers is 93 as [10 Hz – 94 MHz] when a locking bandwidth of 10 kHz is used.

We also found that the timing jitter level of mode-locked fiber lasers can vary significantly depending on the mode-locking conditions, which is set by the input polarization states (by adjusting intra-cavity wave-plates), even at the same intra-cavity dispersion and similar optical spectra. Fig. 6 shows such a case: Even the optical spectra look similar between cases (a) and (b), the jitter spectra can differ more than 10 dB in magnitude. This observation is also consistent with previous measurements in [1], and underlines the importance of BOC-based jitter monitoring to achieve the lowest possible timing jitter performance from mode-locked fiber lasers.

We also experimentally examined the impact of intra-cavity filtering to the high-frequency timing jitter. Recently, there has been an interest in intra-cavity filtering in fiber lasers for shortening pulsewidth [25]. Further, intra-cavity filtering may be used as a restoring force for the indirectly coupled timing noise at the low offset frequency, which has been used for reduction in timing jitter [30]. In our experiment, we inserted a 30-nm bandpass filter (centered at 1040 nm) in both Yb-fiber laser cavities and measured the timing jitter at different mode-locking conditions by adjusting intra-cavity wave-plates. Fig. 7 shows some of the measurement results. Similar to the case without filters (as shown in Fig. 6), depending on the wave-plates setting, the jitter spectra change significantly and we could not observe lower jitter by using an intra-cavity filtering in the high offset frequency regime (e.g., >10 MHz). It is not clear from which offset frequency the intra-cavity filtering can lower the jitter as observed in [30]. Due to the limited locking bandwidth for the BOC method (~2 kHz), we could not resolve the jitter in the lower offset frequency in this work. It will require further study to carefully examine the impact of intra-cavity filtering for lower jitter in the sub-20-as timing jitter mode-locked fiber lasers.

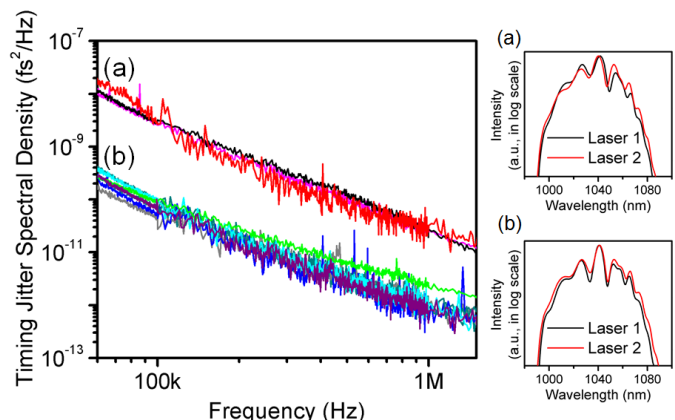

Fig. 6. Timing jitter spectra with different mode-locking conditions show that the magnitude of jitter can vary more than 10 dB even at the same intra-cavity dispersion condition.

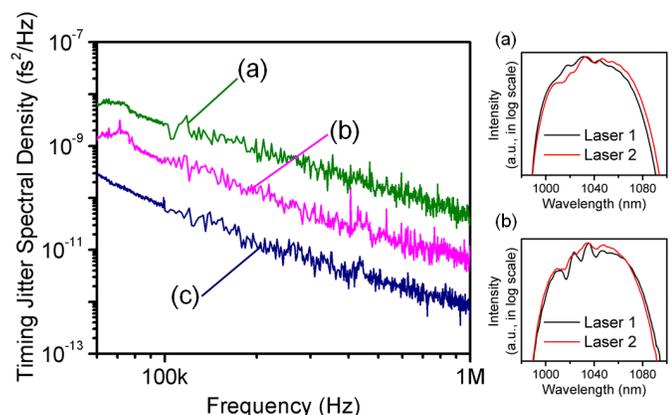

Fig. 7. (a), (b) Timing jitter spectra for different optical spectra when an intra-cavity filtering is applied. (c) The minimum timing jitter spectrum without an intra-cavity filtering.

### B. Timing Jitter Analysis

Based on the soliton perturbation theory-based noise theory of mode-locked lasers [8-11], we analyze the measured timing jitter spectral density as shown in Fig. 8. Curve (a) shows the measured timing jitter spectral density. For >1.5 MHz offset frequency, as already discussed in Section III.A, the measured jitter is limited by the shot-noise-limited noise floor of BOC (curve (b)). From 100 kHz to 1.5 MHz offset frequency range, the measured jitter follows -20 dB/dec slope, which indicates the random walk characteristics of quantum-limited timing jitter. This jitter level can be fit with the calculated quantum-limited timing jitter directly originated from the amplified spontaneous emission (ASE) noise at zero dispersion, as shown by curve (c). The used laser parameters for curve (c) are excess noise factor of 2, FWHM pulsewidth of 40 fs, intra-cavity pulse energy of 2.2 nJ, and gain bandwidth of 45 nm.

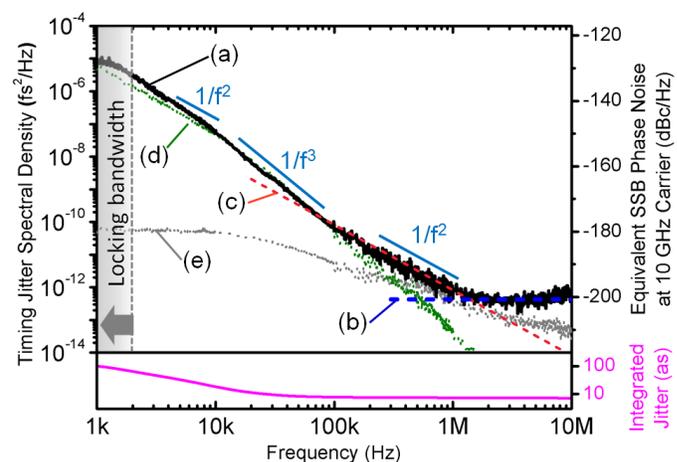

Fig. 8. Timing jitter spectral density analysis. (a) Measured timing jitter. (b) Shot-noise-limited BOC measurement noise floor. (c) Calculated ASE noise originated quantum-limited jitter. (d) Calculated RIN-coupled jitter by self-steepening, based on the direct measurement of RIN-to-jitter coupling coefficient. (e) Calculated RIN-coupled jitter by Kramers-Kronig relationship.

Below the 100-kHz offset frequency, the measured jitter spectrum diverges from the quantum-limited jitter (curve (c)) and increases more rapidly as moving toward the lower offset frequency. This deviation from the quantum-limited jitter may be originated from the RIN-to-jitter conversion via, for example, the self-steepening effect in the laser oscillator [10]. In order to confirm that the origin of this jitter is the coupling from the RIN, we performed a separate measurement of the RIN-to-jitter coupling coefficient, using a method similar to the one shown in [3]: the coupling coefficient transfer function between the RIN and the timing jitter is measured when the pump diode laser is modulated in the 10 kHz – 100 kHz range. Using the measured RIN-to-jitter coupling coefficient ($1.6 \times 10^{13}$) and the measured RIN (as shown in Fig. 1(c)), the jitter spectral density could be expressed as $S_{SS}(f) = 1.6 \times 10^{13}$ RIN$(f)/f^2$ [fs²/Hz], shown by curve (d). It fits very well with the measured jitter below 100 kHz offset frequency and can explain the measured -30 dB/dec slope in the 10 kHz – 100 kHz range. One interesting finding is that the coupling

TABLE I
COMPARISON OF TIMING JITTER PERFORMANCE OF SUB-FEMTOSECOND TIMING JITTER MODE-LOCKED SOLID-STATE AND FIBER LASERS

| Ref | Gain medium (round-trip length) | Mode-locking method | Repetition rate | Integrated timing jitter and integration bandwidth | Measurement resolution in the integration bandwidth |
|---|---|---|---|---|---|
| [1] | Yb-fiber (22 cm) | NPR | 80 MHz | 175 as [10 kHz – 40 MHz] | 20 as |
| [2] | Er-fiber (1.21 m) | NPR | 77.6 MHz | 70 as [10 kHz – 40 MHz] | 24 as |
| [3] | Ti:sapphire (4.56 mm) | KLM | 82.6 MHz | ∼ 10 as [10 kHz – 41.3 MHz] ∼ 125 as [2 kHz – 41.3 MHz] | 8 as |
| [4] | Cr:LiSAF (11 mm) | SESAM | 100 MHz | 30 as [10 kHz – 50 MHz] | ∼ 30 as |
| [5] | Er:Yb-glass (4 mm) | SESAM | 100 MHz | 83 as [10 kHz – 50 MHz] | ∼ 35 as |
| [6] | Er-fiber (79 cm) | ft-CNT (all-fiber) | 78.7 MHz | 490 as [10 kHz – 39.4 MHz] | ∼ 42 as |
| [7] | Yb:KYW (4 mm) | CNT-coated mirror | 1.13 GHz | 700 as [17.5 kHz – 10 MHz] | 410 as |
| This work | Yb-fiber (22 cm) | NPR (short non-gain fiber) | 188 MHz | 14.3 as [10 kHz – 94 MHz] 79.2 as [2 kHz – 94 MHz] | 6.8 as |

* NPR, nonlinear polarization rotation; KLM, Kerr-lens mode-locking; SESAM, semiconductor saturable absorber mirror; ft, fiber tapering; CNT, carbon nanotube.

coefficient is much smaller than the expectation based on the soliton perturbation theory. Assuming the nonlinear phase shift in our fiber laser of >1 rad per round-trip, the self-steepening effect is expected to contribute about 20 dB higher level than our measurement when using analytic formula in [10]. We believe that this damped self-steepening effect and reduced RIN-to-jitter coupling coefficient might be caused by more complex pulse evolution in the stretched-pulse fiber laser (with higher linear and nonlinear chirp and stronger pulse breathing) compared to the solid-state lasers such as Ti:sapphire lasers [3] and Cr:LiSAF lasers [4]. More theoretical and experimental investigation will be necessary to exactly understand this reduced contribution of self-steepening to the timing jitter in stretched-pulse fiber lasers in the future. Finally, curve (e) shows the expected RIN-coupled jitter by the Kramers-Kronig relationship. Provided the measurement is not limited by the BOC resolution, the jitter might be limited by this contribution in the high offset frequency above 1 MHz.

## C. Comparison of Timing Jitter Performance with Other Ultralow-Jitter Mode-Locked Lasers

Table 1 summarizes the performance of sub-femtosecond level timing jitter mode-locked solid-state and fiber lasers demonstrated recently (in 2011-2013). The lowest timing jitter demonstrated so far was from the KLM Ti:sapphire laser [3]: ∼10 as (∼125 as) integrated jitter from 10 kHz (2 kHz) to 41.3 MHz (Nyquist frequency). When 30 kHz locking bandwidth is used, the integrated jitter from 100 Hz to 40 MHz was 13 as. As explained in Section I. Introduction, the shorter pulsewidth (<10 fs), higher pulse energy and simpler pulse dynamics enables the KLM Ti:sapphire laser as the lowest jitter pulse train source. Now, with the short non-gain fiber approach, the mode-locked fiber laser can also generate similar level of ∼10 as high-frequency timing jitter optical pulse train (14.3 as from 10 kHz to 94 MHz in this work).

When comparing the origins of the demonstrated ∼10 as high-frequency jitter of Ti:sapphire lasers [3] and Yb-fiber lasers (in this work), they are slightly different despite similar numbers. Above 1 MHz offset frequency, both measured jitter spectra are limited by the BOC measurement resolution. In the 100 kHz – 1 MHz range, the Yb-fiber laser is limited by the

ASE quantum-limited jitter whereas the Ti:sapphire laser is limited by the technical noise (RIN-to-jitter coupling). In the 10 kHz - 100 kHz range, both lasers have a diverging jitter spectra (with a slope higher than -20 dB/dec) and the Yb-fiber laser has ∼10 dB higher level than the Ti:sapphire laser. In the lower offset frequency below ∼4 kHz, the Ti:sapphire laser has much more rapid increase in jitter spectrum than the Yb-fiber laser, which may be due to more mechanical vibration sensitivity and pump noise. As a result, when the jitter integrated from 2 kHz is compared, the integrated jitter of Yb-fiber laser is lower than that of the Ti:sapphire laser. It can be concluded that although the ASE quantum-limited jitter for the Ti:sapphire laser may be much lower than that of the Yb-fiber laser, due to the technical noise sources (such as the RIN-to-jitter coupling in the oscillator), the timing jitter performance of fiber lasers can be similar to or better than the Ti:sapphire lasers.

## IV. CONCLUSION

In summary, we showed the most recent advancement in femtosecond mode-locked fiber lasers as the ultralow jitter optical pulse train source, reaching the sub-20-as-level high-frequency timing jitter from free-running Yb-fiber oscillators for the first time. The demonstrated jitter performance is already similar to that of the best Ti:sapphire lasers. Due to much reduced cost, easier implementation and better robustness compared to the solid-state lasers, we anticipate the demonstrated fiber lasers can find many applications that require ultimate timing precision.

Further reduction of the non-gain fiber section, by using a special WDM-collimator [31] for example, the jitter performance may be further improved. The study on the impact of intra-cavity filtering at stretched-pulse, self-similar, and all-normal dispersion regimes [26,32] is another topic of study that requires further investigation. Although the Yb-fiber laser was used as the demonstration in this work, we believe the same approach can be used in other types of lasers, for example, in the Er-fiber lasers at the telecom wavelength. Finally, we have an interest in the simultaneous characterization and optimization of timing jitter, carrier-envelope offset phase noise and RIN in mode-locked fiber lasers to the ultimate limits by combining the approaches in [13] and [27].